\documentclass[fleqn,12pt]{article}

\newcommand{\AmS}{{\protect\the\textfont2
  A\kern-.1667em\lower.5ex\hbox{M}\kern-.125emS}}

\title{Inverse beta decay reaction in $^{232}$Th and $^{233}$U fission antineutrino flux}

\author{G. Domogatski$^1$, V. Kopeikin$^2$, L. Mikaelyan$^2$, V. Sinev$^2$ \\
\\
$^{1}$Institute for Nuclear Research RAS, Moscow, \\
$^{2}$Russian Research Center ``Kurchatov Institute"}
\date{} 

\begin{document}

\maketitle


\begin{abstract}
Energy spectra of antineutrinos coming from $^{232}$Th and $^{233}$U neutron-induced 
fission are 
calculated, relevant inverse beta decay $\bar{{\nu}_e}+p \rightarrow n + e^{+}$ positron 
spectra and total cross sections are found. This study is stimulated by a hypothesis 
that a self-sustained nuclear chain reaction is burning at the center of the Earth 
(``Georeactor"). The Georeactor, according to the author of this idea, provides energy 
necessary to 
sustain the Earth's magnetic field. The Georeactor's nuclear fuel is $^{235}$U and, probably, 
$^{232}$Th and $^{233}$U. 
Results of present study may appear to be useful in future experiments aimed to test the 
Georector hypothesis and to 
estimate its fuel components as a part of developments in geophysics and astrophysics 
based on observations of low energy antineutrinos in Nature.
\end{abstract}

\section*{Introduction}

A revolutionary progress in antineutrino ($\bar{{\nu}_e}$) detection technique demonstrated by KamLAND Collaboration 
[1] opens a real opportunities to study $\bar{{\nu}_e}$ fluxes of natural origin and obtain information on their sources, 
which is otherwise inaccessible. 

The Baksan Underground Observatory of the Institute for Nuclear Research RAS (BNO), as discussed 
in [2], is one of the most promising sites to build a massive antineutrino scintillation spectrometer 
for these studies. The research program can include:
\begin{itemize}
\item Study of Thorium and Uranium concentrations in the Earth by means of detection of 
$\bar{{\nu}_e}$ -s coming from beta decay of their daughter products (``Geoneutrinos"). 
This problem was first considered in 60-ies [3] and is intensively discussed presently [4].
\item Estimation of frequency of gravitational collapses in the Universe by detection of 
isotropic $\bar{{\nu}_e}$ flux (1984, [5]).
\end{itemize}

As suggested by R. Raghavan (arXiv:hep-ex/0208038) and discussed in [2], the same spectrometer 
can be used to test the hypothesis [6] that self-sustaining nuclear chain reaction is burning 
at the center of the Earth (''Georeactor"). The main component of the nuclear fuel in the 
hypothetical Georeactor is $^{235}$U, while $^{232}$Th and $^{233}$U fission probably 
also contribute to the reactor power. In this paper we present calculated energy spectra 
of antineutrinos coming from $^{232}$Th and $^{233}$U neutron-induced fission and relevant inverse 
beta-decay positron spectra and total cross sections. Results of present study may appear 
to be useful in future experiments aimed to test the Georector hypothesis and to estimate 
its fuel components as a part of developments in geophysics and astrophysics based on 
observations of low energy antineutrinos in Nature.

Below we shortly describe method used to calculate the $\bar{{\nu}_e}$ energy spectra 
generated by $^{232}$Th and $^{233}$U fission fragments, present relevant $\bar{{\nu}_e}$ and 
positron spectra of the detection reaction 

\begin{equation}
\bar{{\nu}_e}+p \rightarrow n + e^{+}
\end{equation}
 
and compare them with known $^{235}$U spectrum.

\section{Antineutrino and positron spectrum}

1.1 For each of the considered nucleus the $\bar{{\nu}_e}$ energy spectrum 
${\rho}^i_{calc}(E)$ was found by summation of $\sim$550 beta decay spectra of individual 
fission fragments weighted with their fission yields. The fission yields were taken from 
compilation [7], for decay schemes library accumulated in Kurchatov Institute was used.

1.2 Previously similar calculations for $^{235}$U, $^{239}$Pu, $^{241}$Pu and $^{238}$U were 
performed many times by Kurchatov Institute neutrino group and by other authors (a short 
list of references see e.g. in [8]). These studies show the following features:

a) Antineutrino spectra for different fissile isotopes can differ significantly one from the 
other;

b) Uncertainties of calculated spectra are large because of poor knowledge of decay schemes 
for many short-lived fission fragments;

c) The ratios of calculated spectra for different fissile isotopes 
${\rho}^i_{calc}(E)/{\rho}^k_{calc}(E)$ are found with 
considerably lower uncertainties than uncertainties in each of the spectrum 
${\rho}^i_{calc}(E)$ and ${\rho}^k_{calc}(E)$.

We use point c) to correct $\bar{{\nu}_e}$ spectra calculated for $^{232}$Th and $^{233}$U fission.
The following correction procedure has been adopted. We use ''true'' ${\rho}^i_{ILL}(E)$  
$\bar{{\nu}_e}$ spectra) for 
$^{235}$U, $^{239}$Pu and $^{241}$Pu found at ILL by a reconstruction of measured fission 
beta-spectra for each of the fissile isotope 
[9] and find three energy dependent ratios $K_{i}(E)$ of calculated ${\rho}^i_{calc}$ and 
${\rho}^i_{ILL}$ spectra:

\begin{equation}
K_{i}(E)={\rho}^i_{calc}(E)/{\rho}^i_{ILL}(E)
\end{equation}
An averaged over $^{235}$U, $^{239}$Pu and $^{241}$Pu ratio $K(E)$ (Fig. 1) is used to find corrected 
spectra ${\rho}^i_{corr}(E)$:

\begin{equation}
{\rho}^i_{corr}(E) = {\rho}^i_{calc}(E)/K(E),
\end{equation}
$i$ = 2, 3 in Eq. (3) indicates $^{232}$Th and $^{233}$U.

1.3 Found corrected $\bar{{\nu}_e}$ energy spectra for $^{232}$Th and $^{233}$U together with 
the $^{235}$U ILL spectrum are presented in Table 1. Uncertainties include uncertainties 
of the correction procedure 
(i.e. uncertainty in the factor $K(E)$, Fig. 1) and proper uncertainty in ILL spectra. 
Estimated resulting 
uncertainty in ${\rho}^i_{corr}(E)$ is (5-10)\% (68\% CL) in the (1.80-6.0) MeV $\bar{{\nu}_e}$ energy 
region and is increasing to (15-20)\% in the (6-8) MeV energy range. 

Reaction (1) positron spectra $S(E_{vis})$ for $^{232}$Th, $^{235}$U and $^{233}$U 
(Fig. 2a) are presented vs positron energy absorbed in the scintillator

\begin{equation}
E_{vis} \approx E - 1.80 + 1.02 \approx E - 0.8,
\end{equation}

(E - is the energy of the incoming $\bar{{\nu}_e}$, 1.02 MeV is the energy of positron annihilation quanta, 1.80 MeV 
is the reaction threshold).

Total reaction (1) cross sections for $^{232}$Th, $^{235}$U and $^{233}$U fission $\bar{{\nu}_e}$ are presented in Table 2.

Presented data show systematic variation from harder spectra and higher cross sections to softer spectra and smaller 
cross sections in the sequence $^{232}$Th $\rightarrow$ $^{235}$U $\rightarrow$ $^{233}$U. 
This could be expected from the 
neutron/proton contents in the considered nuclei.

\section*{Conclusions}

Energy spectra of antineutrinos coming from $^{232}$Th and $^{233}$U neutron-induced fission are calculated, relevant inverse 
beta decay $\bar{{\nu}_e}  +  p \rightarrow e^{+} + n$ positron spectra and total cross sections are found. Results 
of present study may appear to be useful in future experiments aimed to test the Georector hypothesis and to estimate 
its fuel components as a part of developments in geophysics and astrophysics based on observations of low energy 
antineutrinos in Nature.

\section*{Acknowledgments }

This study is supported by RFBR grant 03-02-16055 and Russian Federation 
President's grant 1246.2003.2.

\vspace{5 cm}

\begin{table}[htb]
\caption{$^{232}$Th, $^{233}$U and $^{235}$U fission antineutrino spectra
(1/MeV $\cdot$ fiss.)}
\label{table}
\vspace{10pt}
\begin{tabular}{c|l|l|l}
\hline
E, MeV & $^{235}$U$^{*)}$ & $^{233}$U & $^{232}$Th \\
\hline
1.75 & $-$ & 1.27 & 1.82 \\
2 & 1.3 & 1.08 & 1.61 \\
2.5 & 0.9 & 0.675 & 1.13 \\
3 & 0.637 & 0.443 & 0.812 \\
3.5 &0.437 & 0.290 & 0.587 \\
4 & 0.283 & 0.177 & 0.405 \\
4.5 & 0.172 & 0.992 (-1) & 0.268 \\
5 & 0.105 & 0.564 (-1) & 0.176 \\
5.5 & 0.617 (-1) & 0.314 (-1) & 0.114 \\
6 & 0.370 (-1) & 0.159 (-1) & 0.672 (-1) \\
6.5 & 0.203 (-1) & 0.778 (-2) & 0.372 (-1) \\
7 & 0.105 (-1) & 0.374 (-2) & 0.201 (-1) \\
7.5 & 0.429 (-2) & 0.137 (-2) & 0.861 (-2) \\
8 & 0.136 (-2) & 0.403 (-3) & 0.272 (-1) \\
\hline
\end{tabular}\\[2pt]
{\small $^{*)}$ The ILL $^{235}$U spectrum [9].}\\
\end{table}

\begin{table}[htb]
\caption{Inverse beta decay total cross sections ${\sigma}_f (10^{-43}$ cm$^2$/fiss.) for 
$^{232}$Th, $^{233}$U and $^{235}$U}
\label{table}
\vspace{10pt}
\begin{tabular}{c|c|c}
\hline
$^{235}$U$^{*)}$ & $^{233}$U & $^{232}$Th \\
\hline
6.39$\pm$ 2.7\% & 3.87$\pm$ 10\% & 9.70$\pm$ 10\% \\
\hline
\end{tabular}\\[2pt]
{\small $^{*)}$ Found using the ILL antineutrino spectrum [9].}\\
\end{table}

\end{document}